\documentclass[12pt]{article}
\usepackage{fullpage,psfig,amsfonts,latexsym,amsmath,epsf,latexsym,amssymb}
 
\newtheorem{theorem}{Theorem}
\newtheorem{lemma}{Lemma}

\newcommand{\R}[0]{{\mathbb{R}}}

\def\C{{\mathbb{C}}}

\def\R{{\mathbb{R}}}

\def\N{{\mathbb{N}}}

\newcommand{\cH}{{\cal H}}

\newcommand{\onemat}[0]{{\mathbf 1}}

\newcommand{\proof}[1]{{\bf Proof.} #1 \hfill $\Box$\vspace{0.5cm}}

\newcommand{\ket}[1]{|#1\rangle}
\newcommand{\bra}[1]{\langle #1|}

\title{{\bf Ergodic quantum computing}}

\author{Dominik Janzing and Pawel Wocjan\thanks{email: \{janzing,wocjan\}@ira.uka.de}\\[1ex] 
{\small IAKS Prof. Beth, Arbeitsgruppe Quantum Computing,} \\
{\small  Universit{\"a}t Karlsruhe,
Am Fasanengarten 5,} \\ {\small 76\,131 Karlsruhe, Germany}}

\date{{\small June 30, 2004}}

\begin{document}

\maketitle

\abstract{We propose a (theoretical\, ;-) model for quantum
computation where the result can be read out from the time average of
the Hamiltonian dynamics of a $2$-dimensional crystal on a cylinder.
The Hamiltonian is a spatially local interaction among Wigner-Seitz
cells containing $6$ qubits. The quantum circuit that is simulated is
specified by the initialization of program qubits. As in Margolus'
Hamiltonian cellular automaton (implementing classical circuits), a
propagating wave in a clock register controls asynchronously the
application of the gates. However, in our approach all required
initializations are basis states. After a while the synchronizing wave
is essentially spread around the whole crystal.  The circuit is
designed such that the result is available with probability about
$1/4$ despite of the completely undefined computation step.  This
model reduces quantum computing to preparing basis states for some
qubits, waiting, and measuring in the computational basis. Even though
it may be unlikely to find our specific Hamiltonian in real solids, it
is possible that also more natural interactions allow ergodic quantum
computing.}

\section{Introduction}
The question which control operations are necessary to achieve
universal quantum computing is essential for quantum computing
research.  The standard model of quantum computation requires (1)
preparation of basis states, (2) implementation of single and
two-qubit gates and (3) single-qubit measurements in the computational
basis. Meanwhile there are many proposals that reduce or modify the
set of necessary control operations (see e.g.\ \cite{RB00,Benjamin,
Benjamin99,LeuMess,AdiabaticQC}). Common to all those models is that
the program is encoded in a sequence of control operations.

Here we consider a model which requires no control operations during
the computation since the computation is carried out by the autonomous
time evolution of a fixed Hamiltonian.  The idea to consider
theoretical models of computers which consist of a single Hamiltonian
can already be found in \cite{Benioff,Feynman:85,Marg86}.  However,
these models are not explicitly designed for implementing {\it
quantum} algorithms. We start from Margolus' approach since it has the
attractive property that the Hamiltonian is a homogeneous spatially
local interaction between cells of a $2$-dimensional lattice and is
therefore ``relatively close'' to interactions in crystals. Margolus'
Hamiltonian implements the dynamics of a classically universal
cellular automaton (CA).  In his two-dimensional model the front of a
spin wave propagates in one direction over the surface and controls
the updating of the cells.  Even though there is no globally
controlled clocking of the updates his local synchronization ensures
that each cell is not updated until all relevant neighbors are already
updated. In the Margolus scheme the computer is always in a
superposition of many computation steps.  At the beginning one has to
prepare the wave front such that it mainly propagates in the forward
direction.  Such a state is not a computational basis state.  We found
it intriguing to use only basis states. Our goal was to reduce the
required control operations to the absolute minimum: input of the
initial state, the writing of the program and the readout of the
classical output.  The basis states we start with consist of
components propagating forward and components propagating backward.
Our circuit is designed such that even the backward computation leads
to the correct result.  When the time average of an appropriate
initial state subjected to the Hamiltonian dynamics is measured one
obtains the correct result with high probability.  The state tells us
whether the result has to be rejected. Hence one may consider the
procedure as a  Las Vegas algorithm. Our
Hamiltonian is a sum of operators which act on $10$ qubits in contrast
to the $2$-dimensional Margolus cellular automaton which needs
interactions between $8$ qubits for universal {\it classical}
computation.  In \cite{PawelPSPACE} one finds a $4$-local Hamiltonian
where the time average of a single qubit encodes the answer of a
PSPACE hard problem.  But the Hamiltonian has to be constructed for
the specific PSPACE problem. The Hamiltonian is not homogeneous and is
not appropriate for universal computation.

The structure of the paper is as follows.  In Section~\ref{Universal}
we choose a set of four $2$-qubit gates which is universal for quantum
computing. In Section~\ref{Constr} we construct a $4$-qubit gate which
includes all these $4$ gates into one controlled gate. This makes the
computer programmable. Then we describe how the synchronization scheme
of Margolus is used: A wave front of a clock register propagating
around the cylinder ensures that the programmable gates are applied in
correct time order. This propagation is done by the evolution of an
appropriate Hamiltonian.  In Section~\ref{Symmetry} we describe the
symmetry of the crystal by the crystallographic concept of
Wigner-Seitz cells.  In Section~\ref{Mix} we prove that the time
average leads to the correct result. The readout of this result is
explicitly described in Section~\ref{Readout}.  In
Section~\ref{PSPACE} we briefly show that ergodic quantum computing
can in principle solve all problems in polynomial space for all
problems where usual quantum algorithms need only polynomial space. At
first sight, this seems to be in contradiction to the fact that time
steps of usual algorithms are translated to spatial propagation (as in
\cite{RB00}).

\section{Universal set of gates}

\label{Universal}

We recall \cite{Kit96} that the following types of gates are
sufficient for universal quantum computation. Let $(\C^2)^{\otimes n}$
be the state space of a quantum register. Then we consider the
following two-qubit and single-qubit gates which are assumed to be
available for every pair of qubits or every single qubit,
respectively:

\begin{enumerate}

\item The Hadamard gate on a single qubit:
\[
H:=\frac{1}{\sqrt{2}}
\left(
\begin{array}{cc} 
1 & 1 \\ 1 & -1 
\end{array}
\right)
\]

\item The controlled-phase gate 
\[
\Gamma(\sigma^{1/2}_z)=|1\rangle \langle 1|\otimes 
 \frac{1}{\sqrt{2}}\left(\begin{array}{cc} 1 & 0 \\ 0 & i \end{array}\right)
+|0\rangle \langle 0|\otimes \onemat 
\]
where $|0\rangle,|1\rangle$ are the canonical basis states of $\C^2$
and $\onemat$  is the identity. 
\end{enumerate}
Note that an exact implementation of the SWAP gate is
possible. Therefore, without losing universality, we allow the
application of controlled phase gates only on adjacent qubits.

We assume that gates acting on disjoint sets qubits can be applied at
the same time step. We call such a time step a layer of the quantum
circuit. The {\it depth} of the quantum circuit is the number of time
steps.

\begin{figure}
\centerline{
\epsfxsize0.5\textwidth\epsfbox[0 0 554 437]{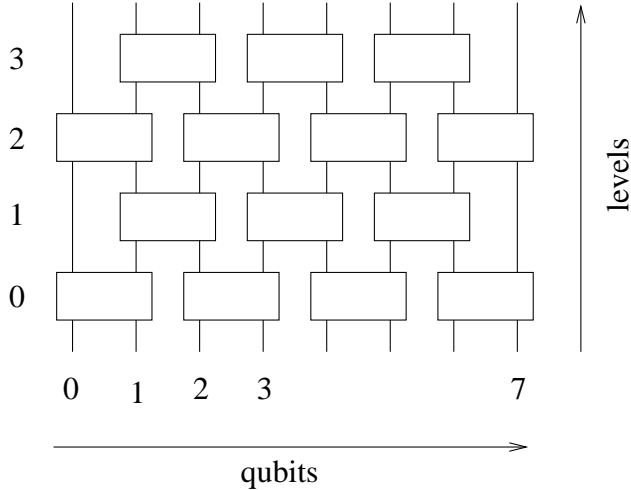}
}
\caption{{\small Decomposition of arbitrary quantum circuits into layers of
two-qubit gates $U_{00},\ldots,U_{11}$ acting on adjacent
pairs.}}\label{fig:layerstructure}
\end{figure}

For reasons that shall be clear later we consider circuits $U$ which
have a special layer structure (see
Fig.~\ref{fig:layerstructure}). Each time step consists of several
gates with the following restrictions:
\begin{itemize}
\item In even time-steps we allow only two-qubit gates acting on
the qubit pairs $(k,k+1)$ with even $k$.
\item In odd time-steps we have only two-qubits gates on $(k,k+1)$
with odd $k$.
\end{itemize}
In this scheme we distinguish formally among four $2$-qubit gates:
\begin{equation}\label{eq:4gates}
U_{00}:=\onemat \otimes \onemat\,,\quad
U_{01}:=\onemat \otimes H\,,\quad
U_{10}:=H \otimes \onemat\,,\quad
U_{11}:=\Lambda(\sigma_z^{1/2})\,.
\end{equation}
 
Using these gates, we construct a circuit $U$ with the following
properties: Let $f:\{0,1\}^n \rightarrow \{0,1\}^m$ be the function we
would like to compute. The unitary $U$ acts on the input, the output
register, and some ancilla register and computes $f$ in the sense
\[
U\big(
|x\rangle \otimes |y \rangle \otimes |0\dots 0\rangle
\big)
=
|x\rangle \otimes |y \oplus f(x)\rangle \otimes |0\dots 0\rangle\,,
\]
where $\oplus$ denotes the bitwise XOR. 
By construction, we have
\[
U^2
\big(
|x\rangle \otimes |y \rangle \otimes |0\dots 0\rangle
\big)
=
|x\rangle \otimes |y \rangle \otimes |0\dots 0\rangle\,.
\]
Without loss of generality, we may assume that $f(x)\neq 0$ for all
inputs $x$ by extending $f$ with an additional bit which is always
$1$.

Note that there are quantum algorithms where $f(x)$ is only computed
probabilistically. We will neglect this fact since it is irrelevant
for the principles of our construction and would make the discussion
unnecessarily technical.

\section{Constructing the crystal Hamiltonian}

\label{Constr}

Usually a quantum circuit is considered as a sequence of
gates. However, the usual way of drawing it (like in
Fig.~\ref{fig:layerstructure}) suggests spatial propagation. Now we
consider quantum circuits where quantum information is really
spatially propagated and the time-axis is represented by the second
dimension.

\begin{figure}
\centerline{
\psfig{file=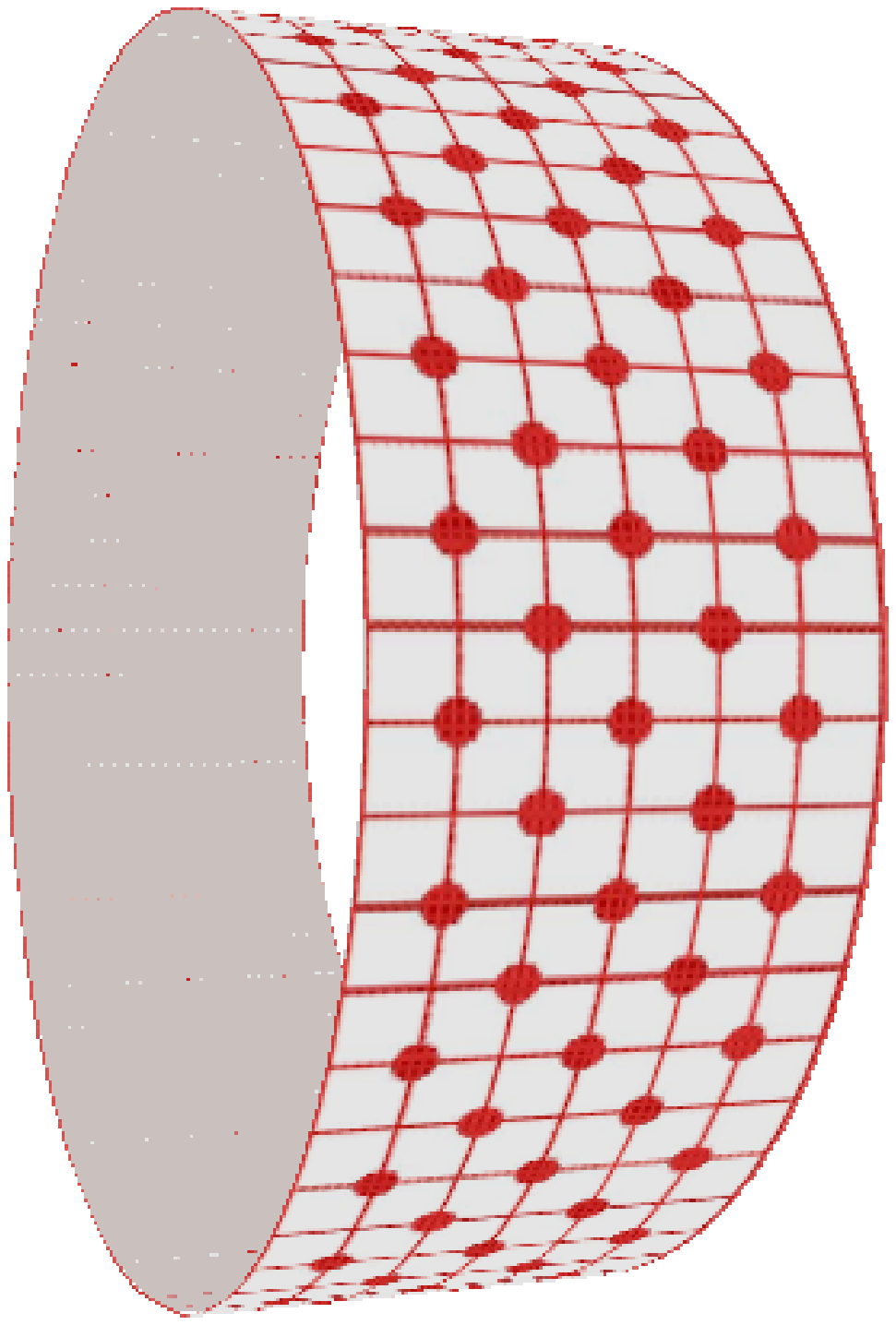,width=5.75cm}
\hspace{1cm}
\psfig{file=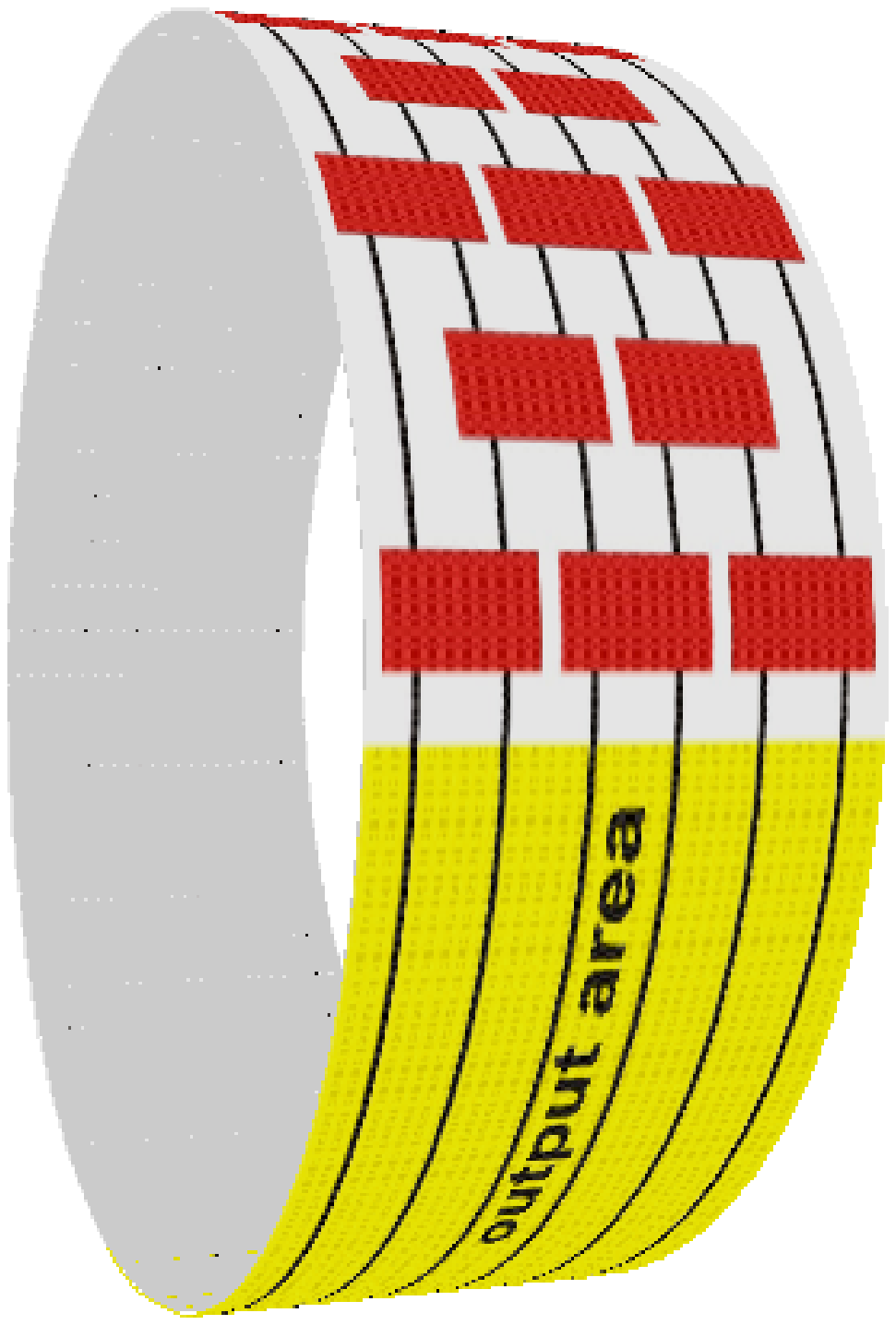,width=5.75cm}
}
\caption{{\small (Left) Cylindric crystal consisting of $c\times h$ cells. A pair of
program qubits is located at the red points. The lines indicate the
boundaries of a cell. (Right) The circuit wrapped around the cylinder. Every 
time when a two-qubit gate is applied the information of both qubits is 
propagated one row upwards. The output region consists only of trivial 
gates, i.e. the information is only propagated.}}
\label{Cyl}
\end{figure}

Our circuit is wrapped around a cylinder. The cylinder is covered by
$c\times h$ squares (``cells'') of equal size. We have $h$ (for ``height'') 
columns
and $c$ (for ``circumference'') 
rows. We need $c>2h$ for reasons which will be clear in Section~\ref{Readout}.
The columns correspond to the qubits of the 
original circuit and the rows to its time steps (see fig.\ref{Cyl}).

Each cell $(j,k)$ contains a data qubit. They form the data space
\[
\cH_D:=(\C^2)^{\otimes ch}\,.
\]
In the $j$-th time step we apply all gates of layer $j$. A gate of the
original circuit acting on the qubit pair $(k,k+1)$ in level $j$
translates to a gates acting on data qubits in cells
$(j,k),(j,k+1),(j+1,k),(j+1,k+1)$. It applies the original two-qubit
gate to the qubits in row $j$ and propagates the information to row
$j+1$. Furthermore, the vertices between those $4$ cells contain two
program qubits which specify which one of the two-qubit gates in
eq.~(\ref{eq:4gates}) should be applied.  Explicitly, there are two
qubits between cell $(j,k)$ and $(j+1,k+1)$ if both $k$ and $j$ are
even or both are odd (see Fig.~\ref{Cyl}).  For each vertex with
program qubits we define the gate
\begin{equation}\label{eq:ourUniversalGate}
V:=W \sum_{l,m \in \{0,1\}\times \{0,1\}} P_{lm} \otimes U_{lm}\,.
\end{equation}
where $P_{jm}:=|jm\rangle \langle lm|$ projects onto the state
$|lm\rangle$ of the two-qubit program register at a certain vertex.
$W$ is the swap gate which exchanges the state of the qubit pairs
$(j,k)$ and $(j+1,k)$ and the pairs $(j,k+1)$ and $(j+1,k+1)$.

\begin{figure}
\centerline{
\epsfxsize0.3\textwidth\epsfbox[0 0 459 175]{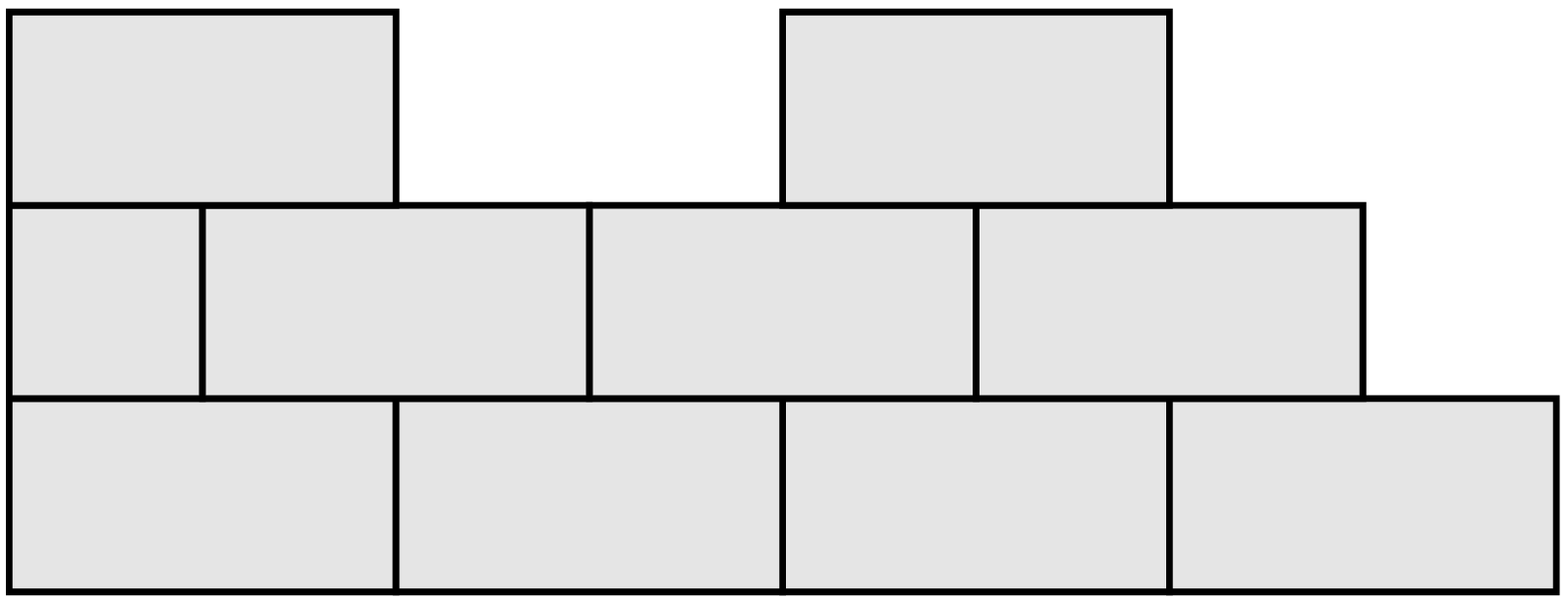}\hspace{2cm}
\epsfxsize0.3\textwidth\epsfbox[0 0 459 175]{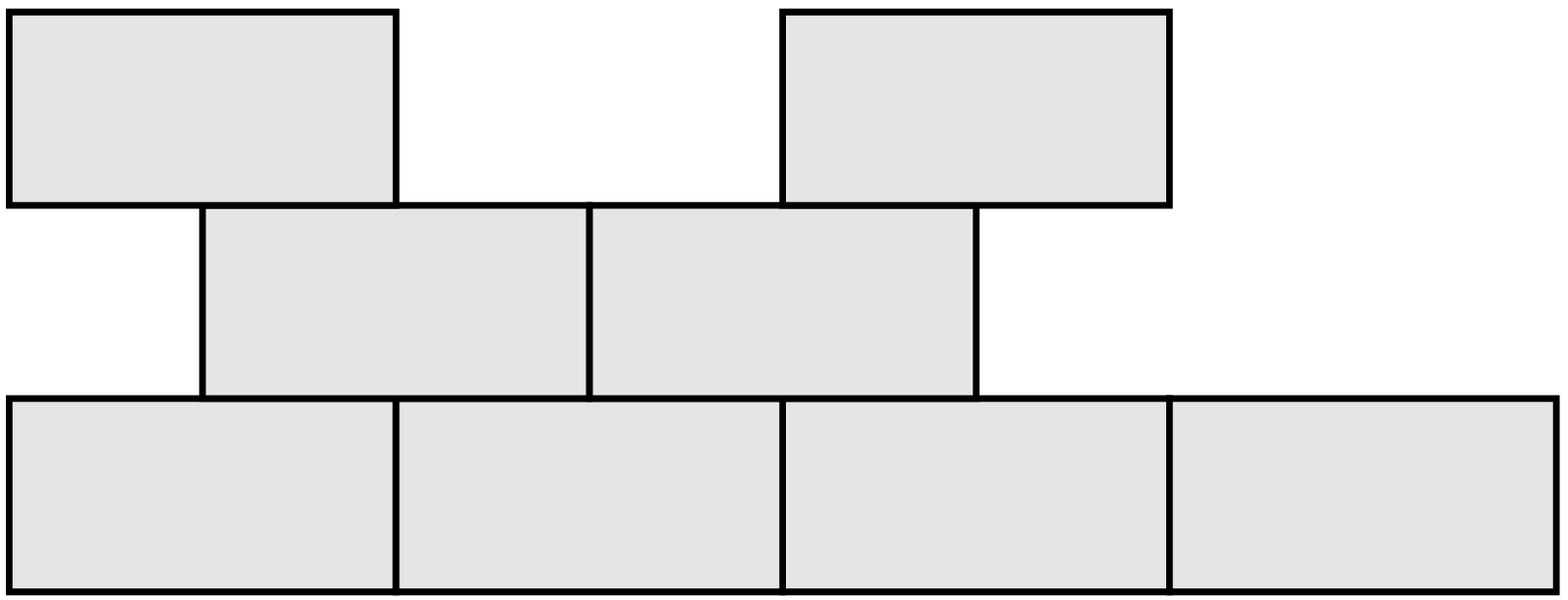}
}
\caption{{\small Correct (left) and incorrect (right) walls. Putting a brick
at position $k,j$ corresponds to carrying out the gate in level $j$
acting on the qubit pair $(j,k)$ and $(j,k+1)$. }}
\label{fig:wall}
\end{figure}

This makes our system programmable and will be essential for achieving
our goal to construct a universal Hamiltonian which can simulate all
circuits. We will only write a program on some part of the cylinder
because we need the other part as output region (see
Section~\ref{Readout}).  As we have already stated, a computation
would consist of applying all gates in row $j$ in the $j$-th step.
However, this requirement is unnecessarily strong. Actually, the only
rule is that each gate in row $j$ can only be applied if both gates in
row $j-1$ which contribute to its input have been applied. These
synchronization rules can be visualized by building walls with bricks
(see fig.~\ref{fig:wall}). The synchronization conditions mean
intuitively that incorrect walls are not allowed. In order to make
this analogy perfect we introduce dummy single qubit gates at the
boundaries of odd rows.

We would like to construct a Hamiltonian such that its autonomous
time-evolution corresponds to a computation which respects these
synchronization rules. Margolus \cite{Margolus:90} solved this problem
by introducing clock qubits as follows.

Each cell contains a clock qubit. Let $\cH_C=(\C^2)^{\otimes ch}$ be
the Hilbert space of all these clock qubits. Define the operator
\[ 
G_{j,k} :=
\begin{array}{ccc}
a^\dagger & \otimes & a^\dagger \\
\otimes   &         & \otimes  \\
a         & \otimes & a
\end{array}\,,
\]
where the annihilation operators $a$ act on the qubits $(j,k)$ and
$(j,k+1)$ and the creation operators $a^\dagger$ act on the qubits
$(j+1,k)$ and $(j+1,k+1)$. These operator $G_{j,k}$ propagate two
$1$'s in the qubits $(j,k)$ and $(j,k+1)$ one row upwards
\[
\begin{array}{|cc|}
\hline
0 & 0 \\
1 & 1 \\
\hline
\end{array}
\quad
\mapsto
\quad
\begin{array}{|cc|}
\hline
1 & 1 \\
0 & 0 \\
\hline
\end{array}\,,
\]
where the left lower corner of the cell is at position $(j,k)$. All
other configurations are mapped onto the zero vector. Now we define
the operator
\[
G:=\sum_{j,k} G_{j,k} 
\]
where $j=0,\ldots,c-1$ and
\begin{eqnarray}
k=
\left\{
\begin{array}{ll}
0,2,\ldots,h-2 & \mbox{ for $j$ even} \\
1,3,\ldots,h-3 & \mbox{ for $j$ odd}
\end{array}
\right.\,.
\end{eqnarray}
In contrast to Margolus we do not consider a cyclic system in both
axis but only cyclic in one direction. This is because we think that a
crystal with $2$-dimensional torus symmetry seems less realistic.

At the boundary we define a family of operators which act on only two
adjacent cells: For each odd $j$ we set
\begin{equation}
G_{j,-1} := \frac{1}{\sqrt{2}}\,
\begin{array}{c}
a^\dagger \\
\otimes \\
a
\end{array}\,,\quad\quad
G_{j,h-1} := \frac{1}{\sqrt{2}}\,
\begin{array}{c}
a^\dagger \\
\otimes \\
a
\end{array}\,,
\end{equation}
where the annihilation operators act on the qubits $(j,0)$ and
$(j,h-1)$ and the creation operators act on the qubits $(j+1,0)$ and
$(j+1,h-1)$, respectively. These operators propagate a $1$ in the
qubits $(j,0)$ and $(j,h-1)$, respectively, one row upwards
\[
\begin{array}{|c|}
\hline
0 \\
1 \\
\hline
\end{array}
\quad
\mapsto
\quad
\begin{array}{|c|}
\hline
1 \\
0 \\
\hline
\end{array}\,,
\]
where the left lower corner of the rectangle is at position $(j,k)$
with $j$ odd and $k=0,h-1$. All other configurations are mapped onto
the zero vector. Now we include these operators in the operator $G$.

\begin{figure}
\centerline{
\epsfxsize0.3\textwidth\epsfbox[0 0 459 175]{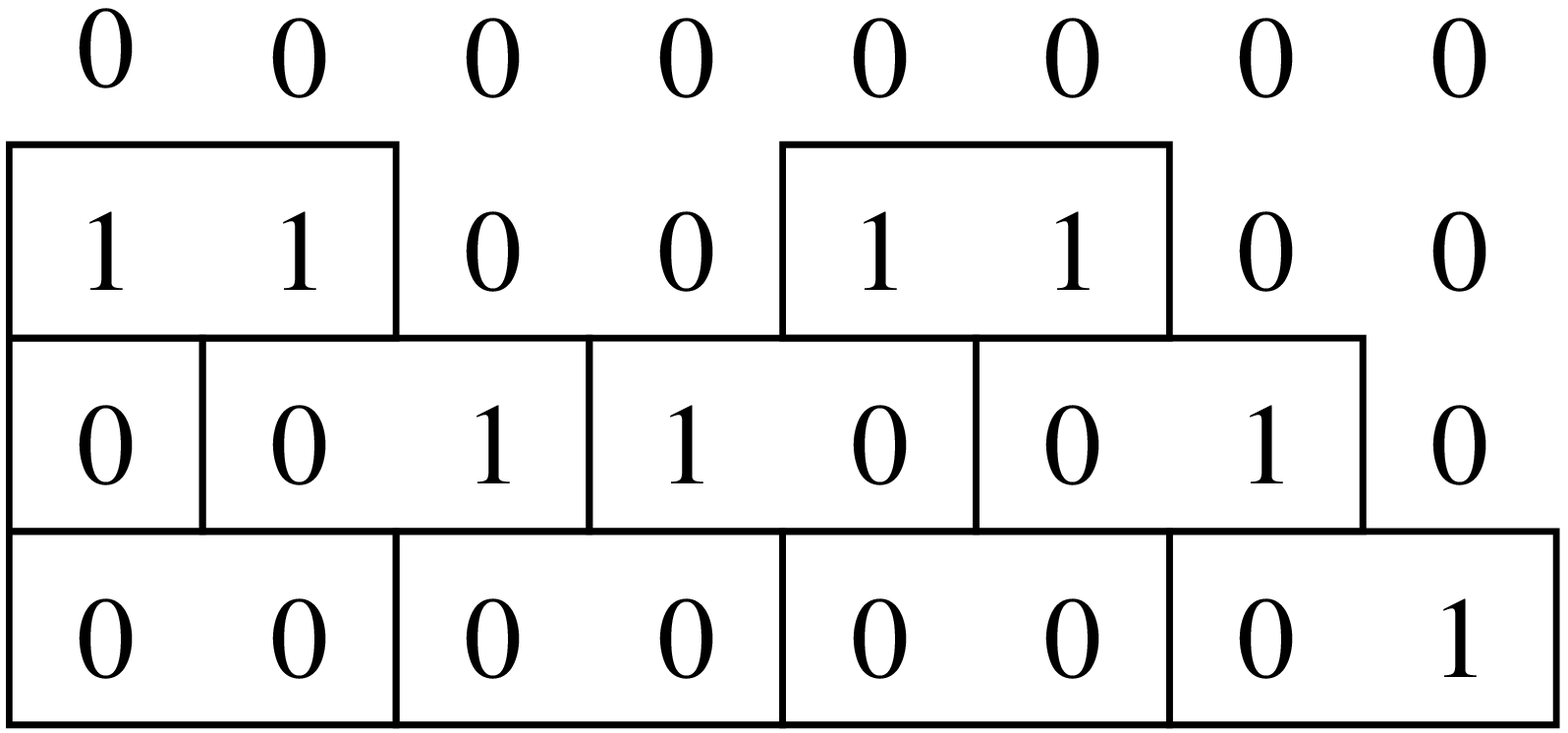}
}
\caption{{\small Visualization of the basis states of $\tilde{\cH}_C$ (the
space of allowed clock states) as brick walls.}}
\label{fig:visuCorrect}
\end{figure}

Now we define a $G$-invariant subspace $\cH_C$, interpreted as the
space of correct synchronizations.  Intuitively, it is spanned by the
set of all basis states corresponding to correct walls. The position
of the uppermost brick in each column is denoted by symbols $1$ as in
fig.~\ref{fig:visuCorrect}.

\begin{lemma}[Synchronization space]\label{lem:synchronSpace}${}$\\
Let $\tilde{\cH}_C$ be the space spanned by those basis vectors
$|a\rangle$, where $a$ is a $0$-$1$-matrix of size $c\times h$
satisfying the following conditions:
\begin{enumerate}
\item[(1)] Each column contains a single $1$, the remaining entries are all
$0$.
\item[(2)] Let $j_k$ be the index of the symbol $1$ in column $k$. Then for
the indices of any two adjacent columns $k$ and $k+1$ we have
\[
|j_k - j_{k+1} | \le 1\,.
\]
\item[(3)] If $k+j_k$ is even then 
\[
j_{k} \ge j_{k+1}\,.
\]
If $k+j_k$ is odd then 
\[
j_{k} \le j_{k+1}\,.
\]
\end{enumerate}
Then $\tilde{\cH}_C$ is $G$-invariant.
\end{lemma}
\proof{Let $a$ be any configuration satisfying the above
conditions.

\begin{itemize}
\item[(1)] Applying $G_{j,k}$ to $\ket{a}$ does not lead to the zero
vector iff the symbol $1$ is at position $j$ in the adjacent columns
$k$ and $k+1$, i.\,e., $j_k=j_{k+1}=j$. If this is the case, then
$G_{k,j}$ propagates both $1$'s one position upward. Therefore, the
configuration $G_{k,j}\ket{a}$ still fulfills condition (1).

\item[(2)] Assume first that $a$ is a configuration with $j_k>j_{k+1}$
for some $k$. Since $a$ satisfies condition (2) we know that
$j_k=j_{k+1}+1$. The only operators which act on qubit $(k,j_k)$ are
$G_{j_k,k}$ and $G_{j_k-1,k-1}$. The $G_{j_k,k}$ operator vanishes
when applied to $\ket{a}$ because the symbol $1$ is at position
$j_k-1$ in column $k+1$ and not at position $j_k$ which would be
required for a non-trivial action of $G_{j,k}$. The operator
$G_{j_k-1,k-1}$ vanishes because the $1$ is at position $j_k$ in
column $k$ and not at position $j_k-1$ as would be required for a
non-trivial action of $G_{j_k-1,k-1}$. This situation is shown on the
left in fig.~\ref{fig:situations}.

\begin{figure}
\centerline{
\epsfxsize0.15\textwidth\epsfbox[0 0 184 170]{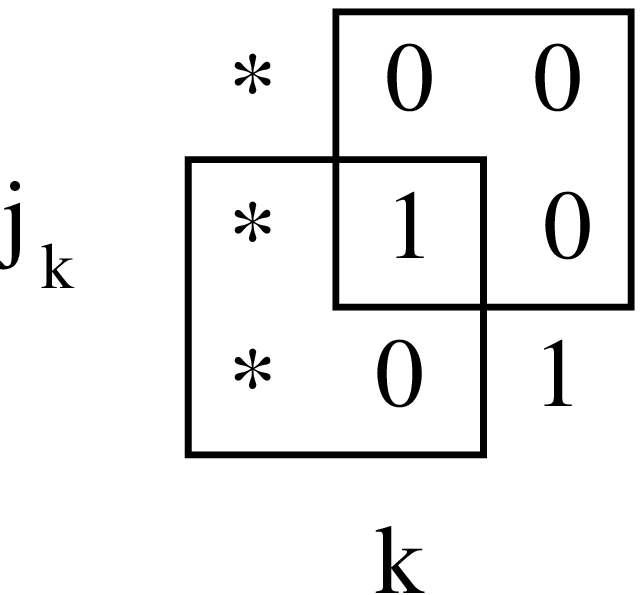}\hspace{3cm}
\epsfxsize0.15\textwidth\epsfbox[0 0 184 170]{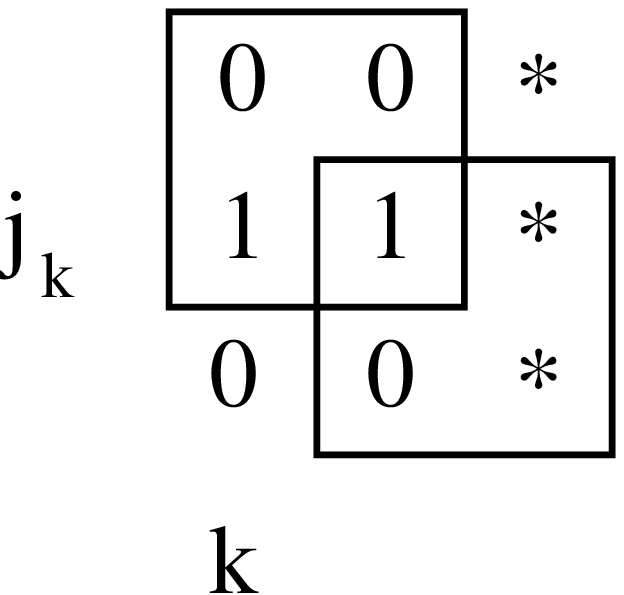}
}
\caption{{\small Left: the application of the operators corresponding to the 
squares annihilate the state. Right: The application of the operator
which corresponds to the upper square propagates both symbols $1$.}}
\label{fig:situations}
\end{figure}

The case $j_k<j_{k+1}$ is proved analogously.

\item[(3)] Assume that $a$ is a configuration with $j_k=j_{k+1}$ for
some pair of adjacent columns $k$ and $k+1$. Let $k+j_k$ be even. In
this case we have to show that every action which increases $j_{k+1}$
also increases $j_k$.

We first consider the case that both $k$ and $j_k$ are even. The only
operators that act on the qubit $(k+1,j_{k+1})=(k+1,j_k)$ are
$G_{k,j_k}$ and $G_{k+1,j_k-1}$. The second operator vanishes because
the symbol $1$ is not at position position $j_k-1$ in the column $k+1$
but at $j_k$. The operator $G_{k,j_k}$ increases $j_k$ and $j_{k+1}$
as claimed. This situation is shown on the right in
fig.~\ref{fig:situations}.

Analogously, we can prove that this is also true if $k$ and $j_k$ are
both odd.  The remaining case is that $k+j_k$ is odd. By using
analogous arguments we can show that every action which increases
$j_k$ also increases $j_{k+1}$.
\end{itemize}
}

\noindent In analogy to Margolus' and Feynman's ideas we define the
forward time operator $F$ by
\[
F:=\sum_{j,k} G_{j,k} \otimes V_{j,k}
\]
where $V_{j,k}$ is the gate $V$ in eq.~(\ref{eq:ourUniversalGate})
acting on the qubits $(j,k)$,$(j,k+1)$,$(j+1,k)$, and $(j+1,k+1)$. For
the operators $G_{j,-1}$ and $G_{j,h-1}$ ($j$ odd) at the boundary we
set $V_{j,-1}:=\onemat$ and $V_{j,h-1}:=\onemat$.

The Hamiltonian is defined as the sum of the forward time operator and
backward time operator
\[
H:=F+F^\dagger\,.
\]
In the sense of \cite{KSV:02} this is a $10$-local interaction since
each operator $G_{j,k}$ acts on $10$ qubits at once. Note that one may
rewrite the interactions as $k$-local terms with $k<10$ by introducing
qudits, i.e., particles with higher dimensional Hilbert
spaces. Therefore the size $10$ does not necessarily mean that this
interaction is unphysical.

To analyze the dynamical evolution we need the feature that $F$ is a
normal operator on the relevant subspace (analog to Margolus'
results). However, since we do not work with cyclic boundary
conditions the proof is a little bit more technical. As noted in
\cite{Biafore} the dynamics of the $1$-dimensional cyclic Margolus
Hamiltonian \cite{Margolus:90} is the quasi-free time evolution of
independent fermions. 

Even though we do not see if the clock dynamics of our Hamiltonian is
also quasi-free, we can prove:

\begin{lemma}
The restriction of  $F$ to the relevant space
\[
\cH:=\tilde{\cH}_C \otimes \cH_D\otimes \cH_P
\]
is normal, i.e., $FF^\dagger=F^\dagger F$.
\end{lemma}
\proof{For an initial state $|\psi\rangle \in \cH$ the operator
 $F F^\dagger|\psi\rangle$ is a
sum of terms of the form
\begin{equation}\label{FF1}
F_{j,k} F^\dagger_{\tilde{j},\tilde{k}}|\psi\rangle\,.
\end{equation}
$F^\dagger F|\psi\rangle$ is a sum of terms of the form
\begin{equation}\label{FF2}
F^\dagger_{\tilde{j},\tilde{k}} F_{j,k} |\psi\rangle\,.
\end{equation}
For $|k-\tilde{k}|\geq 2$ or $|j-\tilde{j}|\geq 2$ the operators
$F^\dagger_{\tilde{j},\tilde{k}}$ and $F_{j,k}$ act on disjoint qubits
and thus commute. Then the products in eq.~(\ref{FF1}) and
eq.~(\ref{FF2}) are equal.

If $|k-\tilde{k}|\leq 1$ and $|j-\tilde{j}|\leq 1$ then it is easily
checked that the product $G^\dagger_{\tilde{j},\tilde{k}}
G_{j,k}|a\rangle$ is only non-zero for $(j,k)=(\tilde{j},\tilde{k})$.

Therefore, it is sufficient to show that
\begin{equation}\label{eq:normality1}
\sum_{(j,k)} F_{j,k} F^\dagger_{j,k} |\psi\rangle =
\sum_{(j,k)} F^\dagger_{j,k} F_{j,k} |\psi\rangle
\end{equation}
in order to prove that $F$ is normal. 
Since the operators $U_{j,k}$ are unitary it is sufficient to show
that
\begin{equation}\label{eq:normality}
\sum_{(j,k)} G_{j,k} G^\dagger_{j,k} |a\rangle =
\sum_{(j,k)} G^\dagger_{j,k} G_{j,k} |a\rangle
\end{equation}
for every allowed clock configuration $a$.
Note that $|a\rangle$ is an
eigenvector of the operators on both sides since each term which does
not vanish is identical to a multiple of the vector $|a\rangle$.
First we consider only the operators $G_{j,k}$ which act on $4$ clock
qubits and not the special operators $G_{j,-1}$ and $G_{j,h-1}$
at the boundaries. In the cyclic model of Margolus, the right-hand term 
in (\ref{eq:normality}) counts the possibilities to go forward and the 
left-hand term the possibilities to go backward.
The fact that both numbers coincide prove normality.
in the non-cyclic case the possibilities to add or remove
half bricks have to be considered separately.

Note that $G_{j,k}|a\rangle$ can only be non-zero if $j=j_k$ (with
$j_k$ defined as in Lemma~\ref{lem:synchronSpace}), i.e., there is the
symbol $1$ in position $j$ in the $k$th column.  Then the term
$G_{j_k,k}|a\rangle$ does not vanish if and only if there is also a
symbol $1$ in position $(j_k,k+1)$.

To formalize these conditions we introduce the variable
$c_k:=(j_k+k)\pmod 2$ indicating whether $j_k+k$ is even or odd.  Due
to the definition of the operators $G_{j,k}$ the term $G_{j_k,k}\ket{a}$
is only non-zero if $c_k=0$. The position of the second symbol $1$
requires $c_{k+1}=1$. Since $j_{k+1}$ can differ from $j_k$ by at most
$1$ the conditions 
\[
c_k=0 \hbox{ and } c_{k+1}=1
\]
are also sufficient that $G_{j_k,k}\ket{a}$ is non-zero.

Similarly, we have $G_{j_k,k} G^\dagger_{j_k,k}|a\rangle=|a\rangle$
whenever
\[
c_k=1 \hbox{ and } c_{k+1}=0\,.
\]
Let $n_{10}$ and $n_{01}$ be the number of occurrences of the patterns
$10$ and $01$ in the string $(c_0,\ldots,c_{h-1})$, respectively. If
$n_{10}=n_{01}$ then the leftmost and the rightmost symbols
coincide. In both cases exactly one of the boundary terms
\[
G^\dagger_{j_0,-1}      G_{j_0,-1}      |a\rangle\,,\quad
G^\dagger_{j_{h-1},h-1} G_{j_{h-1},h-1} |a\rangle
\]
does not vanish and yields the vector $(1/2)|a\rangle$. The same is
true for the terms with the conjugated boundary operators. Hence both
sides of eq.~(\ref{eq:normality}) yield the same vector
$(n+1/2)\ket{a}$.

Note that $n_{10}$ and $n_{01}$ can differ by at most one. This is the
case if and only if the leftmost and the rightmost symbol are
different. If $n_{10}=n_{01}+1$ the leftmost and the rightmost symbols
are $1$ and $0$, respectively. If $n_{01}=n_{10}+1$ they are $0$ and
$1$, respectively. In the first case only the combinations
\[
G^\dagger_{j_0,-1}      G_{j_0,-1}      \ket{a}\,,\quad
G^\dagger_{j_{h-1},h-1} G_{j_{h-1},h-1} \ket{a}
\]
lead to non-zero terms and contribute to the right-hand side of
eq.~(\ref{eq:normality}) with $(1/2)\ket{a}$ each. The conjugated
boundary operators lead both to vanishing terms. This fact compensates
the difference of $1=n_{10}-n_{01}$ in the contribution to the
left-hand and the right-hand side of eq.~(\ref{eq:normality}). The
second case ($n_{01}=n_{10}+1$) is treated analogously.
}

The fact that $F$ is normal helps to understand the dynamical
evolution according to $F+F^\dagger$. In \cite{Margolus:90} this fact
makes it possible to find a conserved quantity interpreted as the
computation speed.  It is given by the operator $V:=(F-F^\dagger)/i$.
Then Feynman and Margolus start with initial states which have a
positive expectation value of the computation speed. Their initial
states are necessarily superpositions of basis states because the
expectation value of $V$ is zero for every basis state of the
clock. Since $|\psi\rangle$ is orthogonal to $F|\psi\rangle$ and
$F^\dagger |\psi\rangle$ we have $\langle \psi| F-F^\dagger
|\psi\rangle=0$ with $|\psi\rangle:=|a\rangle \otimes |\phi\rangle$,
where $|\phi\rangle \in \cH_P\otimes \cH_D$ and $a$ is an allowed
clock configuration.  In our approach, all initial states are basis
states.  Despite these differences, normality of $F$ will be essential
in Section~\ref{Mix} for the ``ergodic theory'' of our Hamiltonian.

\section{Symmetry of the crystal}
\label{Symmetry}
\begin{figure}
\centerline{
\epsfxsize0.5\textwidth\epsfbox[0 0 350 350]{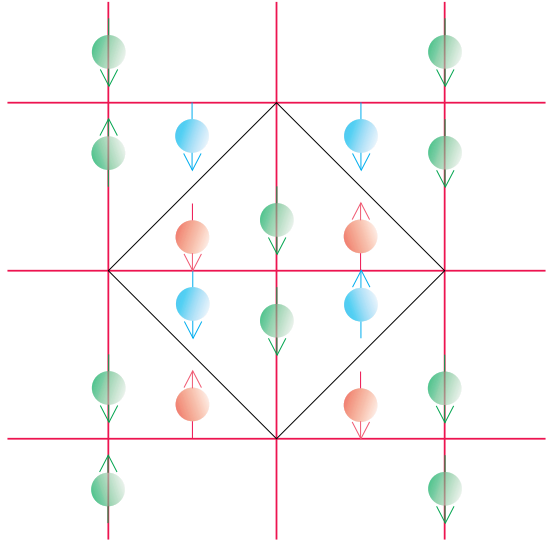}
}
\caption{{\small The black square 
is a Wigner-Seitz cell  containing $6$ qubits. 
The program bits
are green, the clock qubits are red and the data qubits blue. The red grid
indicates the original cells.}}
\label{fig:WScell}
\end{figure}

The symmetry of a crystal can be described by a {\it unit cell} such
that the whole lattice consists of shifted unit cells where the
translations are integer multiples of the lattice vectors.  A usual
way to choose unit cells is given by the so-called {\it Wigner-Seitz
cell} \cite{Zim}. It is constructed as follows.  Consider an arbitrary
point $Q$ in the lattice and consider the set of all points $Q'$ which
are equivalent to $Q$ in the sense that the translation $QQ'$ is a
symmetry operation. Consider the perpendicular bisector of the side
$QQ'$. It divides $\R^2$ into two half-planes containing $Q$ and $Q'$,
respectively. Then the Wigner-Seitz cell (WS cell) is the intersection
of all half planes containing $Q$. Here we choose the position of one
pair of program qubits (i.e., a thick red point in fig.~\ref{Cyl}) as
$Q$.  In the sequel we will refer to our original cells simply as
cells in contrast to the WS cell. A WS cell is a square which has
double area compared to the original cells and is rotated by
$45^\circ$. It covers $4$ adjacent cells such that it contains half of
the area of each. This is depicted in fig.~\ref{fig:WScell}.

We locate the data and clock spins of each cell such that the WS
cell (which is centered around $2$ program qubits) 
contains $2$ clock qubits and $2$ data qubits. Hence, the WS cell
contains $6$ qubits. Each WS cell interacts with  
those adjacent WS cell which have an edge in common.
Note, however, that the interaction among the WS cells
are not pair-interactions between adjacent WS cells because it
contains operators which act on $5$ WS cells at once (note that the 
operators $F_{j,k}$ involve $4$ cells).

Note that the crystal is symmetric under reflections at columns.  due
to the symmetry of the controlled $\Lambda(\sigma_z^{1/2})$-gate.  The
crystal, as we defined it, is not symmetric under reflections at rows.

\section{Mixing properties of the time evolution}

\label{Mix}

Our crystal Hamiltonian $H=F+F^\dagger$ is (on the relevant subspace)
two times the real part of the normal operator $F$. Therefore, $F$ and
$H$ have a common spectral decomposition. The following property of
$F$ is essential:

Let $|\psi\rangle:=|a\rangle \otimes |\phi\rangle \in \cH$ be an
initial state of clock, program, and data register, where $a \in
\{0,1\}^{ch}$ is an allowed clock configuration.  Let $g:=(h+1)c/2$ be
the number of bricks (where half bricks are counted like full bricks)
needed to cover the whole cylindric surface.  Then we have
\begin{equation}\label{Orth}
F^j|\psi\rangle \perp F^k|\psi\rangle
\end{equation}
for all $j\neq k \mod g$. This is easily checked because each state
$F^j|\psi\rangle$ is a superposition of states where ``the wall'' is
enlarged by $j$ bricks.  In order to get the same clock configuration
one needs to add a multiple of $g$ bricks.  Note that the quantum
circuit $U$ (which is encoded in the program register) can be
constructed in such a way that the orthogonality relation~(\ref{Orth})
holds even for all $j \neq k \mod 2g$. Consider, for instance, the
case $j=k+g$.  Project both states in~(\ref{Orth}) onto the subspace
of $\cH$ induced by a definite clock configuration $a$.  On this
subspace, the states of the data register differ by some unitary.
This unitary $U'$ is given by the concatenation of all those gates
which have to be applied in order to go from the clock state $a$ to
$a$ again by winding around the cylinder once.  In other words, $U'$
is obtained by splitting the circuit $U$ and reversing the order of
both parts as follows: Let $U_1$ be given by the sequence of gates
which are applied when the clock wave moves from its initial position
to $a$. Analogously, $U_2$ is given by all gates that are applied when
the clock wave moves from $a$ to the initial position. Then $U'$ is
given by $U':=U_1U_2$ and $U=U_2 U_1$.

If at least one bit of the computed value $f(x)$ is $1$ the
application of $U$ leads always to orthogonal states in the data
register\footnote{If the output part is correctly initialized.}.
Hence the orthogonality relation~(\ref{Orth}) is already satisfied for
$j=g$ and $k=0$. This corresponds to the trivial splitting $U_1=U$ and
$U_2=\onemat$. Unfortunately, the bit flip which occurs on one of the
output bits cannot be implemented by one gate since classical gates
are not available in our setting. Therefore the other splittings may
divide the flip operation into non-classical operations.  In order to
guarantee that also the other splittings lead to orthogonal data
states we may construct $U$ in such way that it flips two bits, one at
the beginning and one at the end. Then either $U_1$ or $U_2$ contain
one complete bit flip.

The following lemma is important for analyzing the ergodic behavior
since it shows that $F$ is essentially a copy of the shift operator
acting on mutually orthogonal spaces:

\begin{lemma}\label{lem:isomorphism}
Let $B$ be a normal operator on a finite-dimensional Hilbert space
$\cH$ and $\ket{\Psi}\in \cH$ such that
\[
B^j \ket{\psi} \perp B^k \ket{\psi} \quad\quad j\neq k \mod N
\]
for some $N\in \N$. Define
\[
\cH_l:={\rm span}_{j\in\N_0} \{ B^{l+j N} \ket{\psi} \}\,,
\]
where $l=0,\ldots,N-1$ and $\hat{\cH}:=\oplus_{l=0}^{N-1} \cH_l$.
Then only the following two cases can occur:
\begin{enumerate}
\item All $\cH_l$ have the same dimension $r$. Then we may identify
$\hat{\cH}$ with $\C^N\otimes \C^r$ such that $\cH_l$ corresponds to
$\ket{l}\otimes \C^r$ and $B$ (restricted to $\hat{\cH}$) has the form
\[
B=S\otimes A + S^\dagger \otimes A^\dagger\,,
\]
where $S$ is the cyclic shift on $\C^N$ and $A$ is some normal matrix
of size $r\times r$.
\item All $\cH_l$ except for $\cH_0$ have the same dimension $r$ and
$\cH_0$ has dimension $r+1$. Then we may identify $\hat{\cH}$ with
$\C\oplus (\C^N \otimes \C^r)$ such that $\cH_0$ corresponds to
$\C\oplus (\ket{0}\otimes \C^r)$ and $\cH_l$ to $\ket{l}\otimes \C^r$
for $l=1,\ldots,N-1$. Furthermore, this identification can be chosen
such that the restriction of $B$ to $\hat{\cH}$ has the form
\[
B=0 \oplus (S\otimes A + S^\dagger \otimes A^\dagger)\,.
\]
\end{enumerate}
\end{lemma}
\proof{Obviously $B$ has the form
\[
\left(
\begin{array}{cccccc}
  0 &     &        &   &         & A_{N-1} \\
A_0 &   0 &        &   &         &         \\
    & A_1 & 0      &   &         &         \\
    &     & \ddots &   & \ddots  &         \\ 
    &     &        &   & A_{N-2} &  0      
\end{array}
\right)\,,
\]
where each $A_l$ maps from $\cH_l$ to $\cH_{l+1 \hbox{ mod } N}$. The
diagonal entries of $B^\dagger B$ are
\[
A_0^\dagger     A_0\,, 
A_1^\dagger     A_1\,,\ldots\,,
A_{N-1}^\dagger A_{N-1}\,.
\]  
The diagonal entries of $B B^\dagger$ are 
\[
A_{N-1} A_{N-1}^\dagger\,, 
A_0     A_0^\dagger\,,\ldots\,,
A_{N-2} A_{N-2}^\dagger\,.
\]  
Since $B$ is normal we conclude
\begin{equation}\label{eq:blockVergleich}
A_{(j+1)\, \hbox{ mod } N}^\dagger\,\, A_{(j+1) \hbox{ mod } N}
= 
A_j A_j^\dagger\,.
\end{equation}
Note that $A_j A_j^\dagger$ and $A_j^\dagger A_j$ have the same rank,
namely $rank(A_j)=rank(A_j^\dagger)$. This shows that all $A_j$ have
the same rank $r$. By definition we have $\cH_{j+1}=A_j \cH_j$ for
$j=0,\dots,N-2$. Therefore, the dimension of $\cH_j$ for
$j=1,\dots,N-1$ is $r$. Only the dimension of $\cH_0$ is not yet
determined. Note that the dimension of $\cH_0$ can not be smaller than
the dimension of $\cH_1$ (since the latter is the image of $\cH_0$).

If $B$ has only trivial kernel in $\hat{\cH}$ then $A_0$ has also
trivial kernel. Then the dimension of $\cH_0$ is also $r$. This
corresponds to the first case.

The following arguments show that we can find a transformation which
changes every $A_j$ to the same matrix $A$. 

Let $B=|B| U$ be the polar decomposition of $B$. Here 
$U$ is unitary and $|B|:=\sqrt{B B^\dagger}$. Note that
$U$ and $|B|$ commute since $B$ is normal. Furthermore, $|B|$ has full
rank and leaves each $\cH_l$ invariant.

We have $U\cH_l=\cH_{(l+1) \mod N}$. Therefore, the power $U^N$ leaves
each subspace $\cH_l$ invariant. Let $X:=\sqrt[N]{U^N}$ in the sense
that $X$ is a function of $U^N$ and not an arbitrary operator $V$ with
$V^N=U^N$.  It also leaves each $\cH_l$ invariant. Define $A$ as the
restriction of $|B |X$ to $\cH_0$. We identify the subspaces $\cH_l$
with $\cH_{(l+1) \mod N}$ with each other via the unitary
transformation $X^{-1} U$. Note that $X$ commutes with $U$. Therefore,
this identification is consistent due to $(X^{-1} U)^N=
X^{-N}U^N=\onemat$. By applying the transformation $(X^{-1} U)^{-1}$
\,\,$l$ times we transport the vector $\ket{\psi}$ to the subspace
$\cH_0$ and obtain $(X U^{-1})^l\ket{\psi}$. By applying $A$ to this
vector and transporting it back from $\cH_{0}$ to $\cH_{l+1}$ we
obtain:
\begin{eqnarray*}
(X^{-1} U)^{l+1} A (X U^{-1})^l \ket{\psi} 
& = & 
(X^{-1} U)^{l+1} |B| X (X U^{-1})^l \ket{\psi} \\ 
& = & 
|B| X X^{-1} U \ket{\psi} \\ 
& = & 
|B| U \ket{\psi} \\ 
& = & 
B\ket{\psi}\,.
\end{eqnarray*}
This shows that our identification of subspaces allows to describe $B$
by the action of the same operator $A$ for each pair $\cH_l$ and
$\cH_{(l+1) \mod N}$. By choosing an arbitrary basis for $\cH_0$ we
may identify all spaces with $\C^r$. This concludes the proof of the
first case.

If $B$ has a non-trivial kernel in $\hat{\cH}$ it is easy to see that
its dimension is $1$. This is due to the fact that the vectors
$B^{l+jN}|\psi\rangle$ are in the image of $B$ for all $j\geq 1$ and
are orthogonal to its kernel. Then we may restrict $B$ to the
orthogonal complement of its kernel and obtain the first case.}

With the isomorphism of Lemma~\ref{lem:isomorphism} we find statements
about the time-average:

\begin{lemma}\label{eklig}
We adopt all notations of Lemma~\ref{lem:isomorphism}. For
$\ket{\psi}\in \cH_0$ define the time-average
$\ket{\psi}\bra{\psi}_T$ by
\[
\ket{\psi}\bra{\psi}_T := \frac{1}{T} \int_{t=0}^T
e^{-i (B+B^\dagger) t} \ket{\psi}\bra{\psi} e^{i (B+B^\dagger) t}\, dt\,.
\]
Let $W$ be the probability measure on $0,\dots,N-1$ defined by
\[
W(l):=tr
\big(
(\ket{l}\bra{l}\otimes\onemat)\, |\psi\rangle \langle \psi |_T
\big)\,.
\]
Let $A=\sum_j a_j Q_j$ be the spectral decomposition of $A$. Assume
that $|\psi\rangle$ lives in the subspace of eigenvalues of $A$ with
large modulus, i.e., $|\psi \rangle \in \sum_j (\onemat\otimes Q_j)
\cH_0$ where $j$ runs over all indices with $|a_j|\geq \epsilon$. 
For 
\[
T\geq \frac{16 N}{\Delta^2\,\, \delta \epsilon} 
\ln \frac{2N}{\Delta}\,.
\]
with $\Delta:=\sin ((\delta/7)^2)$
 the total variation distance between $W$ and the
uniform distribution is at most $\delta$, i.e.,
\[
\frac{1}{2}\sum_l |W(l)-\frac{1}{N}| \leq \delta\,.
\]
\end{lemma}

\proof{We have $\ket{\psi}=\ket{0}\otimes\ket{\kappa}$ with
$\ket{\kappa}\in\C^r$. Hence we have
\[
W(l)=tr\big(
(\ket{l}\bra{l} \otimes \onemat) \,
(\ket{0}\bra{0} \otimes \ket{\kappa}\bra{\kappa})_T 
\big)\,,
\]
where the time average is computed according to the Hamiltonian
$B+B^\dagger=S\otimes A +S^\dagger \otimes A^\dagger$.

The projections $\onemat \otimes Q_j$ commute clearly with
$\ket{l}\bra{l} \otimes \onemat$ and with the Hamiltonian. Hence we
can equivalently consider the time average of the mixture
\[
\sum_j (1\otimes Q_j) |0 \rangle \langle 0|\otimes |\kappa\rangle
\langle \kappa | (1\otimes Q_j)\,.
\]
This is also true if we use $1$-dimensional projections $Q_j$ instead
of the original ones. On the image of each $\onemat \otimes Q_j$ the
time average problem reduces to the following $1$-dimensional
continuous quantum random walk according to the Hamiltonian
\[
\tilde{H}_j = a_j S+ \overline{a}_j S^\dagger\,.
\]
where $S$ is the cyclic shift on $\C^N$. Calculations on the explicit
dynamics can be found in \cite{Gramss} (for $a_j=1$), we are only
interested in time averages. We modify the techniques from \cite{Vazi}
for studying discrete quantum walks to the continuous case.

Let us now consider a fixed index $j$ which is dropped in the sequel.
Then we compute the probability distribution on $0,\dots,N-1$ induced
by the time average of the state $(|0\rangle \langle 0|)_T$ according
the Hamiltonian $\tilde{H}:=aS+\overline{a}S^\dagger$. Let $a=r
e^{i\phi}$ be the polar decomposition of $a$.  Then the eigenvalues of
$\tilde{H}$ are $2 r \cos(\alpha_k)$, where $\alpha_k := \phi + 2\pi
k/N$ for $k=0,\dots,N-1$.  Note that $N$ is even. For simplicity we
consider first the case $r=1/2$ and derive an upper bound on the
mixing time for this case. By rescaling the time we get a general
bound.

Now we consider the system with respect to the Fourier basis.  Then we
denote the eigenvectors of $S$ with eigenvalue $\omega^k:=e^{2\pi i
k/N}$ by $|k\rangle$. The Fourier transform of the original basis
states shall now be denoted by $|e_l\rangle$. The initial state
$|e_0\rangle$ is an equally weighted superposition of all $|k\rangle$,
i.e., the density matrix
\[
\gamma:=\frac{1}{N} \sum_{k,k'=0}^{N-1} |k\rangle \langle k'|\,.
\]
We call eigenvalues and their eigenvectors {\it good} if $|\sin
\alpha_k| \geq \Delta$ with $\Delta =\sin ((\delta/7)^2)$ and denote
the number of good eigenvalues by $N'$.  The length of the interval of
intervals for which $|\sin(\alpha)|\geq \Delta$ for $\alpha\in
[0,2\pi)$ is $4\arcsin\Delta / (2\pi)$. Therefore, we have the
following bound for large $N$:
\begin{equation}\label{deltaklein}
\frac{N-N'}{N} \le \frac{6\arcsin\Delta}{2\pi}\leq
\frac{6 (\delta/7)^2 }{2\pi} \leq (\frac{\delta}{7})^2\,.
\end{equation}
Instead of the superposition of all eigenvectors we consider in the
following an initial vector which is an equally weighted superposition
of only ``good'' eigenvectors:
\[
|e_0\rangle \approx |\beta\rangle :=
\frac{1}{\sqrt{N'}} \sum_{\mbox{good $k$}} |k\rangle\,.
\]
The trace norm distance between the modified density matrix and the 
true initial state is at most
\[
\frac{N-N'}{N}+ 2\sqrt{\frac{N-N'}{N}}\leq 3 \sqrt{\frac{N-N'}{N}}\,,
\]
where the second term in the sum stems from dropping the bad
eigenvalues and the first from rescaling the remaining part.  Using
eq.~(\ref{deltaklein}) it is smaller than
\[
\frac{3\delta}{7}\leq \frac{\delta}{2}\,.
\]

The  time average of the modified state is
\begin{equation}\label{average}
\rho_T :=\frac{1}{T} 
\int_0^T e^{-i \tilde{H} t} \ket{\beta}\bra{\beta} e^{i \tilde{H} t} \,dt=
\frac{1}{N'T} \sum_{k,k'} \int_0^T e^{i(\alpha_{k'}-\alpha_k)t} dt 
|k\rangle \langle k'|\,.
\end{equation}
The distance between two adjacent values $\alpha_k$ is $2\pi/N$.  The
derivative of the cosine is at least $\Delta$ or at most $-\Delta$ for
good eigenvalues. Therefore, for a given $k$ there is at most one $k'$
such that $|\cos(\alpha_k)-\cos(\alpha_{k'})| < \Delta \pi /N$, one in
the interval where the cosine has negative derivative and one in the
other interval with positive derivative.  If we had three values
$\alpha_k,\alpha_{k'},\alpha_{k''}$ such that the distance between
$\cos \alpha_k$ and $\cos \alpha_{k'}$ and between $\cos \alpha_{k'}$
and $\cos \alpha_{k''}$ is less than $\Delta \pi /N$ then we would
have $|\cos \alpha_k-\cos \alpha_{k''}| < \Delta 2\pi/N$. Then we have
at least two vales $\alpha_k$ in the same interval which cannot be
this close to each other due to the assumption on the derivative.

Define projections $Q_p$ for every equivalence class $p$, i.e., $Q_p$
projects onto the span of all $|k\rangle$ with $k\in p$. Note that
these spaces are either $1$-dimensional or $2$-dimensional. We want to
show that the probability distribution
\begin{equation}\label{Pdef}
P(l):=\langle e_l | \rho_T  |e_l\rangle
\end{equation}
is {\it almost} the uniform distribution on the $N$ points
$l=0,\dots,N-1$. We start by showing that the modified distribution
\begin{equation}\label{Rdef}
R(l):=\langle e_l |\sum_p Q_p \rho_T Q_p |e_l\rangle
\end{equation}
is {\it almost} uniform. Explicitly, we have
\begin{equation}\label{RMass}
R(l)=
\frac{1}{N'\, N} 
\sum_{k,k'} (1 + (f(k,k')\, \omega^{(k-k')l})
\end{equation}
where 
\[
f(k,k'):= 
\left\{
\begin{array}{cl}
\frac{1}{T}\int_0^T e^{i (\cos \alpha_{k'} -\cos \alpha_k) t}
dt & \quad\mbox{for $k$ and $k'$ equivalent}\\
0 & \quad\mbox{otherwise}
\end{array}
\right.
\]
and the sum runs over all ordered pairs $(k,k')$ of good indices.

We measure the distance between the probability distributions $P$ and
$R$ by the total variation distance
\[
\|P-R\|:=\frac{1}{2}\sum_l |P(l)-R(l)|\,.
\]
The {\it Diaconis-Shahshahani bound} \cite{Diaconis} estimates the
total variation distance from $R$ to the uniform distribution $U$ by a
sum over the Fourier coefficients of $R$:
\[
\|R-U\|\leq \frac{1}{4}  \sum_{m\neq 0} |\hat{R}(m)|^2 
\]
Note that the first term of eq.~(\ref{RMass}) has only a contribution
to $\hat{R}(0)$.  Hence we have only to consider the second term.  We
obtain
\[
|\hat{R}(m)| \leq 
|\frac{1}{N'\, N} 
\sum_{l=0}^{N-1} \sum_{k,k'} 
\omega^{-lm}  f(k,k') \omega^{l(k-k')}|
\leq 
\sum |f(k,k')|\,,
\]
where the last sum runs over $k,k'$ such that $k-k'=m \mod N$.  There
is at most one equivalent pair $k,k'$ satisfying this condition. The
reason is that one index $k$ is in the region with negative derivative
of the cosine and one index $k'$ in the positive region.  Let $l,l'$
be another equivalent pair where $l$ is in the negative region.  Since
$l$ and $k$ are in the same region we may assume without loss of
generality $l=k+d$ with $d < N/2$.  If $l-l'= k-k' \mod N$ we must
have $l'=l+d \mod N$.  Then $\cos \alpha_l \leq \cos \alpha_k - d
\Delta 2\pi/N$ and $\cos \alpha_{l'} \geq \cos \alpha_{k'} - d \Delta
2\pi/N$.  Hence $l$ and $l'$ cannot be equivalent.  Therefore we find
\[
\sum_{m\neq 0}
|\hat{R}(m)|^2 
=
\frac{1}{N'\, N} \sum_{(k,k')} |f(k,k')|^2 \leq \frac{1}{N}\,.
\]
The last inequality is due the fact that there are at most $N'$
ordered equivalent pairs (of good eigenvalues). This proves
$\|R-U\|\leq 1/N$ which is clearly smaller than $\delta/4$ for
sufficiently large $N$.

Now we consider the total variation distance between $P$ and 
$R$. Using the explicit representation~(\ref{average}) of $\rho_T$ 
and the definitions of $P(l)$ and $R(l)$ in  eq. 
(\ref{Pdef}),(\ref{Rdef}) we have
\[ 
\|P-R\|=
\frac{1}{2N'T}\sum_l \,\sum_{k,k'} 
|\langle e_l |k\rangle|\,| \langle k' |e_l\rangle|\, 
\big|\int_0^T  e^{i(\cos \alpha_k -\cos \alpha_{k'})t} \, dt \big|\,,
\]
where the sum runs over all good  inequivalent ordered pairs  $(k,k')$.
Note that we have $|\langle e_l|k\rangle |=1/\sqrt{N}$.
Due to
\[
|\frac{1}{T}\int_0^T e^{ixt} dt| \leq \frac{2}{T|x|}
\]
we have
\[
\|P-R\|\leq \frac{1}{N'T}\sum_{k,k'}
\frac{1}{|\cos \alpha_k -\cos \alpha_{k'}|}\,.
\]
For fixed value $k$ we divide the inequivalent values
$\cos(\alpha_{k'})$ in classes $m=1,\dots, \lceil 2N/\Delta
\rceil$ such that
\begin{equation}\label{mInt}
\frac{\Delta m }{N} \leq |\cos(\alpha_k)-\cos(\alpha_{k'})|
< \frac{\Delta (m+1)}{N}\,.
\end{equation}
The cosine function is on the interval $[0,2\pi)$
two to one and its derivative has at least modulus $\Delta $
for the good eigenvalues. Therefore we have for a fixed $k$ for every $m$ 
at most two $k'$ such that 
\[
\cos \alpha_{k'} \in [\cos \alpha_k+\frac{\Delta m}{N},\cos \alpha_k
+\frac{\Delta (m+1)}{N}]\,,
\]
for $2N/\Delta \geq m \geq -2N/\Delta$. Hence the inequality
(\ref{mInt}) is at most for $4$ values fulfilled. Therefore have
\begin{eqnarray*}
\|P-R\| &\leq & 
\frac{4N'}{\Delta N' T}  \sum^{2N/\Delta}_{m=1}  
\frac{1}{m} \leq 
\frac{4}{\Delta T}
(\frac{2N}{\Delta})  \ln (\frac{2N}{\Delta})\,.
\end{eqnarray*}
In order to have this term less than $\delta/4$ one has to wait
the time
\[
\frac{32 N}{\Delta^2 \,\,\delta} \ln \frac{2N}{\Delta}\,.
\]
Rescaling the dynamics with the modulus $2 r_j$ of the eigenvalues of
$A$ the time is increased by the factor $1/\min\{2
r_j\}=1/(2\epsilon)$. Then we obtain the time $T$ as stated above.

Putting everything together we obtain
\[
\|W - U\| \le
\frac{\delta}{2} + \|P-R\| + \| R - U \| 
\le \delta\,,
\]
where $\delta/2$ stems from the restriction to good eigenvectors.}

For the initial state vector of our computation we have the problem
that we do not know a priori whether a large component lies
in the kernel of $F$. This is important since this component
remains stationary under the evolution. For the part in the image
we would like to know whether a large component lies in the subspace
of small eigenvalues. This component would require a long mixing time.
To address both problems we need the following two lemmas:

\begin{lemma}\label{Kernel}
Let $B$ be a normal operator on a Hilbert space and $|\psi\rangle$ 
an arbitrary unit vector. Let $\alpha$ be the angle between
$|\psi\rangle$ and $B|\psi\rangle$.
Let $P$ be the projection onto the 
image of $B$. Then
\[
tr(P|\psi\rangle \langle \psi| P) \geq  \cos^2 \alpha \,.
\]
\end{lemma}

\proof{
The projection of $|\psi\rangle$ onto the image of $B$ has at least
the length of the projection of $|\psi\rangle$ onto 
the span of $B|\psi\rangle$ since the latter is a subspace of the image. 
Hence the projection onto the image has at least the length
$\cos \alpha$.
}

In order to estimate  the mixing time we need the following:

\begin{lemma}\label{eigenvalues}
 Let
$L:=\|B|\psi\rangle\|$ be the length of
$B|\psi\rangle$ and $0<\delta < L$. Let $P_\delta$ be the projection onto
all eigenspaces of $F$ with eigenvalues of modulus at least $\delta$.
Then we have
\[
tr(P_\delta |\psi\rangle \langle \psi| P_\delta) \geq  \cos^2 (\alpha -\arcsin(\delta/L))\,,
\]
with $\alpha$ as in Lemma~\ref{Kernel}.
\end{lemma}

\proof{Define the operator $B_\delta:=P_\delta B$. Due to
$\|B_\delta-B\|\leq \delta$ the tip of the vector
$B_\delta|\psi\rangle$ is in an $\delta$-sphere around the tip of
$B|\psi\rangle$.  By elementary geometry, the angle between
$B_\delta|\psi\rangle$ and $|\psi\rangle$ is at least $\alpha
-\arcsin(\delta/L)$.  Since $P_\delta$ is the projection onto the
image of $B_\delta$ we obtain the statement using Lemma~\ref{Kernel}
above.}

\begin{figure}
\centerline{
\epsfxsize0.25\textwidth\epsfbox[0 0 500 500]{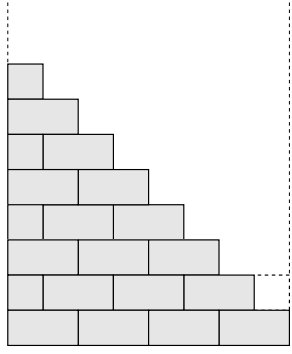}
}
\caption{{\small Initial configuration where the component in the kernel
of $F$ can be estimated. The dashed brick indicates the only possibility
to add a brick.}}\label{InitialWall}
\end{figure}

To use the lemmas above we could use the initial state
$a\in\{0,1\}^{ch}$ of the clock which is indicated by the wall in
Fig~\ref{InitialWall}. The only possibility to add a brick is at the
rightmost position (cell $1,h-1$).  Hence
$F|a\rangle=(1/\sqrt{2})|a'\rangle$ where $a'$ is the new wall with
the additional half brick.  To calculate $F^\dagger|a'\rangle$ note
that $a'$ allows only two ways to remove a brick, namely that one just
added (then $a'$ is mapped to $a$ again) and the upper most brick on
the left ($a'\mapsto a''$).  This means that
\[
F^\dagger F|a\rangle = \frac{1}{2}(|a\rangle \oplus |a''\rangle)\,.
\]
Hence the angle between $F^\dagger F|a\rangle $ and $|a\rangle$ is 
$\pi/4$. The length of $F^\dagger F|a\rangle$ is 
$L=1/\sqrt{2}$. 

Using Lemma  \ref{Kernel} we obtain
\[
tr( P|a\rangle \langle a|P) \geq \cos^2 (\pi/4)= 1/2\,,
\]
and by Lemma \ref{eigenvalues} we have
\[
tr( P_\delta |a\rangle \langle a|P_\delta) \geq \cos^2 (\pi/4 -\arcsin 
(\sqrt{2}\delta )\,.
\]
Note that eigenvalues of $F^\dagger F$ of modulus $\delta$ correspond
to eigenvalues of $F$ with modulus $\sqrt{\delta}$.
If $\tilde{P}_\epsilon$ is the spectral projection of $F$ for eigenvalues
of at least modulus $\epsilon$ we have
\[
tr(\tilde{P}_\epsilon |a\rangle \langle a| \tilde{P}_\epsilon) 
\geq \cos^2 (\pi/4 -\arcsin 
(\sqrt{2}\epsilon^2 ))\,.
\]

We will use the lemmas in this section to estimate the probability of
success of the ergodic quantum algorithm. The idea is as follows.  In
Section~\ref{Readout} we will argue that the correct result can be
found for all states $F^j|\psi\rangle$ (where $|\psi\rangle$ is the
initial state) for all $j$ satisfying a certain condition. This
condition ensures that the circuit $U$ has been applied an odd number
of times on the data qubits. To formalize this we introduce spaces
$\cH_l$ as in lemma~\ref{lem:isomorphism} which are spanned by the
vectors $F^{l+jN}|\psi\rangle$.  Let $Q_l$ be the projection onto
$\cH_l$.  Then for the probability distribution induced by the time
average
\[
W(l)=\frac{1}{T} \int_0^T
 tr( Q_l  e^{i -Ht}|\psi\rangle \langle \psi| e^{iHt})\, dt
\]
we have a lower bound on each $W(l)$: 

\begin{lemma}\label{TimeA} 
There is an initial state of the clock configuration such that
the probability $W(l)$ to find the state in $\cH_l$ after one has
waited 
the time $T$ as in lemma~\ref{eklig}
is at least
\[
W(l)\geq (\frac{1}{N} -2\delta) \cos^2(\pi/4 -\arcsin(\sqrt{2}\epsilon^2))\,.
\]
\end{lemma}

The proof follows immediately from the lemmas of this Section: We
choose the initial clock configuration of
fig.~\ref{InitialWall}. Above we have argued that the probability for
finding a state in the eigenspace with eigenvalues of modulus at least
$\epsilon$ is given by the $\cos^2$-expression on the right.  Given a
state in this subspace we have uniform distribution up to a variation
distance $\delta$.  This yields the factor $1/N - 2\delta$.

In the next section we show for which part of the spaces $\cH_l$ we have
certainly a correct result and how this  promise  is used 
in the readout procedure.

\section{Initialization and Readout}
\label{Readout}

It is clear that the program qubits have to be initialized according
to the simulated quantum circuit. Furthermore we have to initialize
all clock qubits. On the data register we have only to initialize those
qubits which are located in the cells where the initial clock wave
is located.

The readout of the computation result is done as follows.  Here we
assume that the initial state of the clock register is $|a\rangle$
where all symbols $1$ are in row $0$ (In Section~\ref{Mix} we have
also considered another initial configuration which makes is easier to
decide which component of the initial state is in the kernel of the
Hamiltonian. However, the analysis of this Section is technically more
complicated and the output region would have to be enlarged for this
initial configuration). We define an output region which consists of
all cells with column index between $1$ and $m$ where $m$ is any
natural number greater than $2h$.

We choose an arbitrary row in the output region and measure as many
clock qubits of this row as are necessary to find the wave front. If
we have found a clock qubit in state $|1\rangle$ in position $j,k$ the
wave front in row $j+1$ and $j-1$ has to be in one of the columns
$k-1$, $k$, or $k+1$. By this procedure we can localize the whole wave
front. If it is completely localized in the output region we know that
the state of the corresponding logical qubits is either of the states
$U(|x\rangle \otimes |0\dots 0\rangle$) or $|x\rangle \otimes |0\dots
0\rangle$.  Then we can readout the result. We may define $f$ in such
a way that we can decide whether the result is correct or not. In the
following we will give a lower bound on the success probability of the
whole readout procedure.  First we estimate the probability for
finding the wave front in the output region.

The wave front starts in row $0$. States in $\cH_l$ are in general
superpositions of different wave fronts. Note that every such wave
front consists of $l \mod g$ bricks. By elementary geometric arguments
one can check the following statements: First we consider the case
that $l$ is in the interval $0,\dots,g-1$.  A wave front which
consists of more than $h^2/2$ bricks has completely passed row $0$.
Similarly, all row indices of the symbols $1$ can be guaranteed to be
at most $m$ if $l$ is at most $h(m-h)/2$. Therefore, we have at least
$h(m-h)/2-h^2/2=h(m-2h)/2$ spaces $\cH_l$ which are completely in the
output region.  We obtain the same number of spaces $\cH_l$ for
$l=g,\dots,2g-1$.  By these arguments we can easily derive the
following lower bound.  For each space $\cH_l$ we can guarantee at
least the probability $1/N-2\delta$. This yields the following bound.

\begin{lemma}[Probability to find the wave front]\label{lemma:Output}${}$\\
The probability for finding the wavefront completely in the output
region is at least
\[
h(m-2h) (\frac{1}{N} - 2 \delta) s_\epsilon
\,,
\]
where $s_\epsilon$ is the size of the component of the initial state
in the eigenspace of $F$ with eigenvalues of modulus at least
$\epsilon$.
\end{lemma}

However, only the values above in the second interval ensure correct
output.  Since the function $f$ is without loss of generality $1$ on
at least one bit, we can distinguish whether the result has to be
rejected and the experiment has to be repeated.  The probability that
the first experiment succeeds can hence be estimated by dividing the
lower bound of Lemma~\ref{lemma:Output} by two:

\begin{theorem}[One Shot Success probability]\label{Endtheorem}${}$\\
The probability for finding the output region and furthermore
obtaining the correct computation result
is at least 
\[
\frac{h(m-2h)}{2}(\frac{1}{N}-2 \delta) s_\epsilon\,,
\]
with $s_\epsilon$ as in Lemma~\ref{lemma:Output}.
\end{theorem}

Note that we have chosen the initial state of Fig.~\ref{InitialWall}
because we were able to prove a lower bound on the length of the
component in the image of $F$ which yields a good value for
$s_\epsilon$.  Actually, the disadvantage of this initial state is
that the propagation is slow since the wall can grow only at one
point. The more natural initial configuration given by a flat wall
allows propagation in every second column. For this wall, we do not
have a good estimation for the component in the image of $F$ which is
as simple as for the wall in Fig.~\ref{InitialWall}. Nevertheless, we
believe that it is better to start with the flat wave.  If the size of
the output region dominates the size of the circuit (i.e. $c \approx
m$ and $c\gg h$) the quotient $h(m-2h)/N$ tends to $1$ since
$N=c(h+1)$. With small $\epsilon$ the factor in
Theorem~\ref{Endtheorem} is almost $1/2$.  Hence the success
probability tends to $1/4$.

\section{Solving PSPACE problems in crystals of polynomial size}
\label{PSPACE}
It seems to be a general property of our construction that the size of
the crystal necessarily grows linearly with the running time (i.e.,
the depth) of the encoded circuit.  From the  complexity
theoretic point of view, this would have important 
consequences.  Note that the complexity
class PSPACE contains all problems which can be solved using
polynomial space resources \cite{Papa}. The running time of an
algorithm solving a problem in PSPACE may be exponential.  This seems
to imply that the ergodic quantum computer would need exponential
space in contrast to usual models of computation (e.g. 
Turing machines and
Boolean circuits). Now we want to show briefly that even the
ergodic quantum computer can solve all problems in PSPACE in
polynomial space.

The key idea is that even if an algorithm has exponential running time,
it has necessarily (by definition) a polynomial description
of the required sequence of operations. Therefore it is always possible
to construct a circuit $U$ of polynomial depth  such that the
repeated application of $U$ solves the PSPACE problem.

In \cite{PSPACE} we have shown that for every problem in PSPACE
there is a two-gate quantum circuit $U$ of polynomial size which
computes a function $f:\{0,1\}^n \rightarrow \{0,1\}^m$
in the following sense:

\begin{enumerate}

\item
There is a (possibly exponentially large) natural number $r$
such that
\[
U^r (|x\rangle \otimes |y\rangle \otimes |0\dots,0\rangle )
=
|x\rangle \otimes |y\oplus f(x)\rangle \otimes|0 \dots 0\rangle\,,
\]
where $x\in \{0,1\}^n$ is the input string and $y$ is an arbitrary
string in the output register\footnote{The construction in \cite{PSPACE}
is restricted to binary functions. However, the generalization to 
several output qubits is straightforward.}.

\item 
The change of the state of the output register given by 
\[
y\mapsto y\oplus f(x)
\]
occurs for a certain power $s$ of $U$, i.e., for all $U^j$ with $0\leq
j<s $ the output state is still $y$ and for all $U^j$ with $ s \leq j
\leq r-1$ it is already $y\oplus f(x)$.

\end{enumerate}

Furthermore, $r$ and $s$ are known by construction of $U$.
This is possible since there is always an upper bound
on the running time of an algorithm derived from the restricted 
space resources. By introducing idle cycles (counting steps) 
one can guarantee that this bound is exactly attained.
Note that it does not make sense to require that the change of
the output state occurs during the $r$th application  of $U$.
Otherwise $f$ could be computed by a single application of $U^{-1}$.
This is shown by the following argument:

Assume 
\[
U^r(|x\rangle \otimes |0\dots 0\rangle \otimes |y\rangle)
=|x\rangle  \otimes |0\dots 0\rangle \otimes |y \oplus f(x)\rangle
\]
and
\[
U^{r-1}(|x\rangle \otimes |0\dots 0\rangle \otimes |y\rangle)
=|\psi \rangle \otimes |y\rangle\,,
\]
where $|\psi\rangle$ is an appropriate state of ancilla+input register.
Then we have
\begin{eqnarray*}
U^{-1} (|x\rangle  \otimes |0\dots 0\rangle \otimes |y\oplus f(x)\rangle)
&=& U^{-1} U^r (|x\rangle  \otimes |0\dots 0\rangle  \otimes |y\rangle)\\
=|\psi\rangle \otimes |y\rangle
\end{eqnarray*}
This mean that one application of $U^{-1}$ maps $y\oplus f(x)$ onto 
$y$, i.e.,
$0$ is mapped onto $f(x)$. 

The construction of \cite{PSPACE} follows the usual philosophy of
reversible computation \cite{Bennett:73}: The actual computation is
done during the first $r/2$ cycles. Then the result is copied to the
output register with Controlled-Not gates. The only goal of the last
$r/2$ cycles is to undo the computation and restore the initial state.

The ergodic theory in Section \ref{Mix} applies directly to 
PSPACE problems after
substituting $N=2g$ to $N:=2rg$. Furthermore one has to guarantee
the orthogonality condition~(\ref{Orth}) for all $j\neq k \mod 2rg$.
The bit flips which have been explained at the beginning of 
Section~\ref{Mix}  have to be substituted by  incrementing
counter registers. 

The readout is done exactly as in Section~\ref{Readout}.
Given that we have localized the clock wave front in the output
region we have the correct result with probability $1/2$. 
As in  Section~\ref{Readout} we can choose $f$ in such a way that
it indicates whether the result has to be rejected.
Hence the probability of success is not reduced by the fact
that the computation requires more cycles of $U$.

\section{Conclusions}
We have proposed a model of quantum computing which does not require
any control operations during the computation process. The only
required operations are the initialization of basis states and the
readout in the computational basis.

The relevance of this model is two-fold: first it shows that, in
principle, quantum computation can be realized with a small amount of
quantum control. Even though our interaction is rather artificially
constructed, it is a priori not clear that it is unphysical: It
consists of finite range interactions among cells of a crystal which
contain some finite dimensional quantum systems.  This shows that
relatively simple local interactions in homogeneous solid states may
have universal power for quantum computing without external control.
We admit that it seems difficult to decide whether the interactions in
real matter have such properties. However, this may be an interesting
question for future research.

The second aim of this paper concerns the thermodynamics of computation.
As in  \cite{Benioff,Feynman:85,Marg86}
the  computation is performed in an energetically
closed physical system 
with the additional feature that only the preparation of basis states
is required.

It would be desirable to find more simple Hamiltonians which are
universal for ergodic quantum computing. A basis to find them could be
given by simple $1$-dimensional universal quantum cellular automata.

\subsection*{Acknowledgments}
Thanks to Khoder Elzein for drawing the nice cylinders. This work has
been supported by grants of the project ``Kontinuierliche Modelle der
Quanteninformationsverarbeitung'' (Landesstiftung Baden-W\"{u}rttemberg).



\end{document}